\documentclass[eqsecnum,aps,twocolumn,showpacs]{revtex4}

\usepackage{dcolumn}
\usepackage{graphicx}
\usepackage{amsmath}

\begin{document}

\title{Site determination and thermally assisted tunneling in homogenous nucleation}

\author{Jascha Repp$^{1,2}$, Gerhard Meyer$^{1,2}$, Karl-Heinz Rieder$^{2}$, and Per Hyldgaard$^{3}$}

\affiliation{
$^{1}$IBM Research, Zurich Research Laboratory, CH-8803 R\"uschlikon, Switzerland}
\affiliation{
$^{2}$Institut f\"ur Experimentalphysik, Freie Universit\"at Berlin, Arnimallee 14, 
D-14195 Berlin, Germany, }
\affiliation{$^{3}$Department of Applied Physics, Chalmers University of
Technology and G\"oteborgs University, S-41296, G\"oteborg, Sweden}

\date{September 8, 2003}

\begin{abstract}
A combined low-temperature scanning tunneling microscopy 
and density functional theory study on the binding 
and diffusion of copper monomers, dimers, and trimers adsorbed on 
Cu(111) is presented. 
Whereas atoms in trimers are found in fcc sites only, 
monomers as well as atoms in dimers can occupy the 
stable fcc as well as the metastable hcp site.
In fact the dimer fcc-hcp configuration was found to be only 1.3\,meV
less favorable with respect to the fcc-fcc configuration.
This enables a confined intra-cell dimer motion,
which at temperatures below 5\,K is dominated by thermally assisted tunneling.
\end{abstract}

\maketitle
\label{sec:level1}

In recent years important progress has been made in understanding epitaxial growth 
processes on the atomic scale, from both the experimental and the theoretical side. 
Early experiments employed field ion microscopy (FIM), whereas in recent years 
scanning tunneling microscopy (STM)
proved to be a perfect technique 
for this task.

Interestingly, the first steps in the homoepitaxial growth  on fcc (111) noble-metal surfaces
still belong to the unsolved problems. Compared to the more open metal surfaces, 
the behavior of single adatoms is more complicated on (111) surfaces, because adatoms 
can reside in two different binding sites, the so-called fcc and hcp sites. 
Whereas the fcc site continues the ABC layer stacking of the bulk the 
hcp site breaks this sequence.
As the energy difference between the two sites and the related diffusion barriers can be very small, 
the site occupation preference is expected to depend critically on the size of the metal structures, 
i.\,e., it is different for monomers, dimers and trimers 
as was demonstrated experimentally for Ir/Ir(111)\,\cite{Wang90,Busse03}. 

The experimental determination of the fcc/hcp site preference of individual adatoms is a difficult task,
and only two experimental studies exist. 
Wang and Ehrlich determined single adatoms of Ir/Ir(111) to be located on hcp sites \cite{WangEhrlich89,WangEhrlich9192}, whereas 
in the case of Pt/Pt(111) a clear fcc site preference was found\,\cite{Goelzhaeuser96}. 
A general rule predicting the site preference was later developed for metals having a partly filled 
$d$-band\,\cite{Piveteau92,Papadia96} 
that can be applied in both cases. 
Whereas no experimental work can be found in the literature
for systems without a partly filled $d$-shell, several theoretical 
studies exist.
First-principles theoretical work by  Stumpf and Scheffler\,\cite{StumpfScheffler} 
and Feibelman\,\cite{Feibelman92} 
predicted an hcp site preference for individual Al atoms on Al(111). 
Other theoretical studies on copper, silver, and gold
do not find any clear hcp or fcc site preference for a single 
adatom adsorbed on the fcc(111) surface\,\cite{Boisvert95,Kuerpick01}.
Clearly, very subtle contributions of both the electronic structure and 
the local relaxation
have to be taken into account \cite{GiesenIbach}, 
and therefore no general rule is available to predict 
the preferred binding site. Although in homoepitaxy Ir/Ir(111)
is the only known experimental example of an hcp site preference so far, 
several similar systems may well exist.

In this work we determined the adsorption site of Cu metal 
adatoms to be the fcc site, by means of a unique technique of STM diffusion studies  
in combination with atomic manipulation, 
applicable also for systems that cannot be studied by FIM. 
Moreover a particular localized diffusion behavior 
of  Cu dimers has been observed 
experimentally, with the properties as predicted earlier in a theoretical work by 
Bogicevic {\it et al}.\,\cite{Bogicevic98}. 
Despite the large mass of Cu atoms, this localized diffusion
is dominated below 5\,K by thermally assisted tunneling of Cu adatoms.
Copper as a noble metal is 
characterized by an almost complete $d$-band occupation, making
the existing, qualitative theoretical analysis 
of site preference~\cite{Piveteau92,Papadia96} 
indeterminate. 
Instead the experimental results are complemented by an extensive set 
of large-scale {\it ab-initio} density functional theory (DFT) 
calculations of the site preference on Cu(111).

Our experiments were performed with a low-temperature STM
operated at $4-21$\,K.  
The sample was cleaned by 
sputtering and annealing cycles. Bias voltages refer to the sample
voltage with respect to the tip.  
To measure
possible tip effects we determined the hopping
rate for different tunneling parameters
and varied the
tip-adatom interaction time by taking image series with
fixed parameters but for different time intervals.
Using atomic manipulation the monomers and dimers 
were positioned far away from defect sites.
Great care was taken to calibrate the temperature measurements precisely.

Our calculations are
within the generalized gradient approximation (GGA)
using the plane-wave ultrasoft-pseudopotential 
code DACAPO\,\cite{DacapoLB}.
We used slab supercell geometries with 
a vacuum separation $\gtrsim13$\,\AA, and
with a five- (four-)layer slab thickness and
a 3-by-3 (4-by-5) atom in-surface extension to
map the monomer (dimer) adsorption and dynamics.
The two lowermost layers were kept frozen in a bulk-lattice
structure.
The adsorbates and surface atoms 
were relaxed to 0.05\,eV/\AA\ 
using an energy cutoff of 30 Ry, with dipole correction
enabled.
Our DFT calculations should determine the local relaxation,
binding, and vibrations extremely accurately. 
However, {\it ab-initio} DFT can encounter difficulties in resolving 
the energy variation between different adsorption sites if they are 
in the range of only a few meV~\cite{PtPuzzle01}.
Consequently, we have 
tested the robustness of our results and used a carefully controlled 
sampling of $k$-points: a standard 
choice of $2\times2$ with a 0.1\,eV Fermi smearing tested 
against $4\times4$ $k$-point calculations with a 0.025\,eV smearing. 
In addition, we have tested the monomer site-preference result
against calculations using a fully relaxed four-layer 
5-by-4 unit cell.

The diffusion paths of individual adatoms were 
mapped by taking a series  
of consecutive STM images on Cu/Cu(111).
Five series of more than 100 images each 
were taken at different temperatures between 11 and 12\,K,
allowing the determination of the diffusion barrier as 
$E_B=(37\pm 5)$\,meV\,\cite{Repp00}, in good agreement with other 
experiments\,\cite{Wulfhekel96,Knorr02} as well as 
theory\,\cite{Stoltze94,Kuerpick01}.  The mapping of the relative 
movement of one adatom in between subsequent images 
clearly reveals the hexagonal lattice of the substrate 
surface, showing that only one adsorption site is observed.
As pointed out by Wang and Ehrlich\,\cite{WangEhrlich9192}, 
the conventional technique of mapping diffusion paths 
is not applicable for a site determination 
if only one of the sites is occupied.
On the other hand, lateral manipulation on the atomic scale\,\cite{EiglerSchweizer90,Meyer00}
can in general be used as a tool 
to  access also metastable sites by artificial occupation of non-equilibrium 
configurations \cite{Heinrich02}.

\begin{figure}
\centerline{\includegraphics[width=7cm]{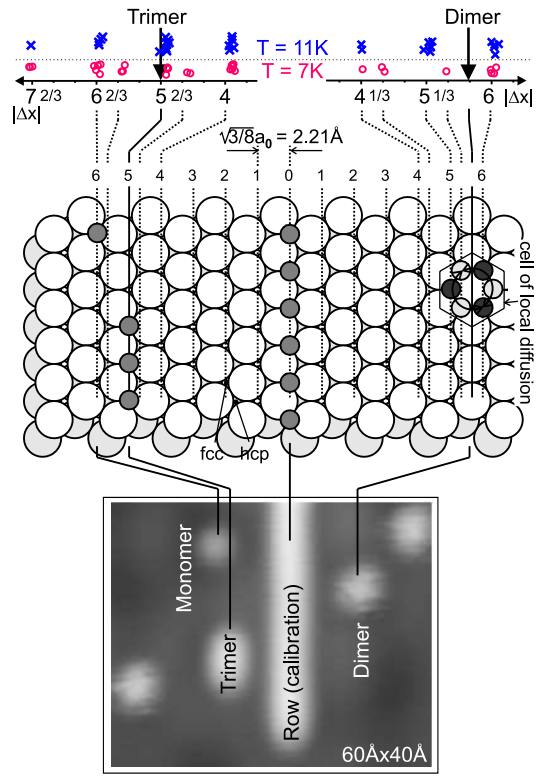}}
\caption[2]{By means of an extended site-mapping technique 
the adsorption site of Cu/Cu(111) 
was determined 
to be the fcc site. Individual adatoms were repeatedly repositioned 
using lateral manipulation at a temperature of 7 and 11\,K. 
One STM image of a series and the corresponding model are shown.
In the topmost part of the figure, the positions of these individual adatoms (circles at 7\,K, crosses at 11\,K) as well as of a dimer and a trimer (arrows) with respect to the row are shown.
}
\label{FigMonDimTri}
\end{figure}
For this type of extension of the mapping technique,
individual adatoms were repeatedly repositioned 
using lateral manipulation at a temperature of 7\,K.
At this temperature the monomers can be positioned in both the fcc and the hcp
sites, making the site determination possible.
If the temperature is increased to $\simeq10$\,K the individual 
adatoms regain their equilibrium position, which can now be identified as the fcc site (see Fig.\,\ref{FigMonDimTri}).

In order to have a point of origin for this extended site-mapping technique, 
the positions were determined 
with respect to a monatomic copper row built up atom by atom.
The row ran parallel to one of the close-packed [$1\bar10$] directions.
The lateral distances between the individual adatoms and the row 
were $|\Delta x|=n\cdot\sqrt{3/8}a_0$, with integer values of $n$ 
for the fcc positions and fractional values of $n$ for the hcp positions.
In addition, this shows that the atoms in the row reside 
in fcc positions in agreement with our theoretical calculations~\cite{TrimerRowDFT}.

The diffusion experiments, in which no 
hcp monomer occupation could be identified, revealed
that the hcp-to-fcc hopping rate is at least 75 times
higher than the fcc-to-hcp hopping rate,
providing a lower bound for the energy difference of the 
two sites of $|\Delta E|\geq kT\cdot\ln(75)=4$\,meV.
Alternatively, the hopping from hcp to fcc sites, which sets in at a 
temperature that is $\Delta T\simeq1.5$\,K lower than the onset for fcc-to-hcp hopping,
allows a rough estimate of the energy difference 
for the two sites as $\Delta E\simeq E_B\cdot\Delta T/T$.
Taking the experimental uncertainties into account, this provides an upper bound 
of $\Delta E\leq8$\,meV,
subject to the assumption that the fcc and hcp diffusion
prefactors differ by no more than a factor of ten (in fact our theory
calculations yield identical prefactors of $\nu_0=1\times 10^{12}$s$^{-1}$ \cite{monovibration}).

This result is consistent with a previous DFT study \cite{Bogicevic00}
and congruent with our DFT calculations 
that yield an fcc preference of about 6\,meV.
An identical energy difference of 6\,meV arises in a 
corresponding DFT study of monomer adsorption on a 
prerelaxed frozen surface, or when we increase the 
$k$-point sampling while lowering the smearing. A similar 
fcc preference arises when, instead, we use a four-layer 
5-by-4 unit cell.
These tests~\cite{Bogicevic00} suggest that the monomer calculations are converged 
with respect to the choice of DFT parameters.
The small monomer-adsorption preference
reflects subtle differences in the interplay 
between the adsorbate-induced density reponse and the exact 
location of subsurface atoms, as we shall discuss further in a 
forthcoming publication.

\begin{figure}
\centerline{\includegraphics[width=6cm]{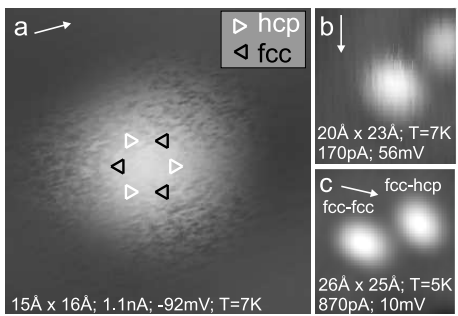}}
\caption[3]{(a) At $T=7$\,K Cu dimers appear unstable in STM images, showing that local diffusion 
takes place during scanning. (b) Long-range interactions 
with a monomer close by can stabilize the dimer.
(c) At a temperature of 5\,K, the dimer diffusion is slow enough to trace individual diffusion steps.
The white arrows indicate the fast scan direction.
The crystal orientation 
corresponds 
to the one 
in Fig.\,\ref{FigMonDimTri}.
}
\label{FigDimer}
\end{figure}
In experiments, dimers do not exhibit any translational diffusion up to a temperature of 21\,K.
Nonetheless they already diffuse locally at a temperature as low as 7\,K,
as can be seen from the unstable appearance of the dimer in the STM image shown in
Fig.\,\ref{FigDimer}(a). 
As predicted by theory\,\cite{Bogicevic98} 
and verified by the site determination demonstrated in Fig.\,\ref{FigMonDimTri}, 
the local diffusion takes 
place in a cell centered around one atom in the topmost layer 
of the substrate surface.
This cell of local diffusion consists of three hcp and three fcc sites 
around the on-top position of the substrate atom (see scheme in Fig.\,\ref{FigMonDimTri}).
In the following we will use the notation 
$ff$, $hh$, $fh$, $fb$, $hb$, and $bb$ for
fcc-fcc, hcp-hcp, fcc-hcp, fcc-bridge, hcp-bridge, and bridge-bridge dimer occupation,
respectively. 

As can be seen in Fig.\,\ref{FigDimer}(b) 
a dimer can be stabilized by long-range 
interactions\,\cite{Repp00,HyldgaardPersson00,Fichthorn00} with 
another monomer positioned close to the dimer.
Moreover, reducing the temperature to below 7\,K,
the dimer intra-cell diffusion becomes so slow 
that it is possible to trace the individual diffusion steps 
by recording a series of STM images.
The stable $ff$ and the metastable $fh$ configuration 
can be observed (see Fig.\,\ref{FigDimer}(c)\,).
The relative probability of observation at 5\,K 
($P_{fh}/P_{ff}\simeq 1/20$)
yields an energy difference of only $\Delta E=-kT\cdot\ln(P_{fh}/P_{ff})=(1.3\pm0.5)$\,meV.
The $hh$ configuration could not be found.

\begin{figure}
\centerline{\includegraphics[width=7cm]{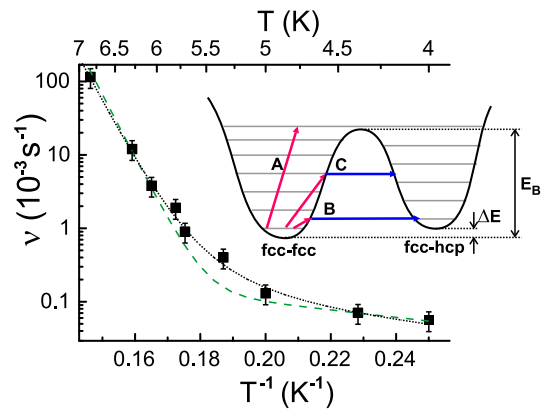}}
\caption[2]{Arrhenius plot of the dimer $ff$ to $fh$ hopping rate.
The hopping rate is determined by direct thermal activation 
(process A) at temperatures above 6\,K and 
by thermally assisted tunneling to the ground state (process B) below 5\,K.
The dashed line represent a fit, taking only these two processes into account. 
At 4.5\,K\,$\leq T\leq6$\,K 
an even better fit (dotted line) is obtained by including
also thermally assisted tunneling for all intermediate energies (processes C).
Tunneling at $U=10$\,mV, $I=10$\,pA excluded 
tip-induced diffusion.}
\label{DimArr}
\end{figure}
An Arrhenius plot of the dimer $ff$ to $fh$ hopping rate (Fig.\,\ref{DimArr})
documents that thermally assisted tunneling \cite{Goran,LauhonHo,Heinrich02}
enables the hopping of the dimers at very low temperatures ($T\leq5$\,K), despite the large mass $m_{\rm Cu}$ of a Cu atom.
In a simple model with a parabolic barrier at the transition state $fb$, characterized by an imaginary
frequency $\omega_{fb}$ through $E=E_B+1/2\cdot m_{\rm Cu}\omega_{fb}^2x^2$, the 
transmission probability due to tunneling for an energy $\epsilon$
is known to be approximately $\exp(-2i\pi(E_B-\epsilon)/\hbar\omega_{fb})$ \cite{HillWheeler}.
In the case of thermally assisted tunneling this probability has to be multiplied by
the probability for thermal activation to the energy 
$\epsilon$ of $\exp(-\epsilon/kT)$. This results in 
a critical temperature $T_C=\hbar\omega_{fb}/2i\pi k$, which marks the limit between
thermally activated hopping and thermally assisted tunneling.
Above this temperature we obtain the conventional Arrhenius behavior 
with the activation energy $E_B$, whereas for the low-temperature regime, 
only thermally assisted tunneling to the ground state is relevant, and
the slope in the Arrhenius plot is determined by 
the energy difference $\Delta E$ of the two ground states.
Only in the temperature region just below $T_C$ do all intermediate 
processes come into play (see fits in Fig.\,\ref{DimArr}).
The fit to the data yields $\nu_0=8\cdot10^{11\pm0.5}$\,s$^{-1}$ and $E_B=18\pm3$\,meV.
Assuming the barrier to be cosine-shaped, $\omega_{fb}$ and thus 
$T_C$ can be estimated from the distance $d=1.47$\,\AA\ between the fcc and the hcp 
sites and the barrier height $E_B$ as $T_C\simeq\sqrt{E_B/2m_{\rm Cu}}\cdot \hbar/ kd=6.0$\,K,
in agreement with experiment. The surprising observation of the tunneling of a heavy Cu atom
can be understood when considering that
the atom not only has a very low barrier 
but also an exceptionally small distance to overcome.
In contrast to the recently reported tunneling of carbon adatoms \cite{Heinrich02}, we do not see any 
influence of descrete vibrational levels of the dimer. This indicates a very strong vibronic 
coupling of the dimer to the substrate, clearly ruling out the possibility of a 
Cu$_2$/Cu(111) quantum rotor \cite{Bogicevic98}. 

\begin{figure}
        \centerline{\includegraphics[width=8cm]{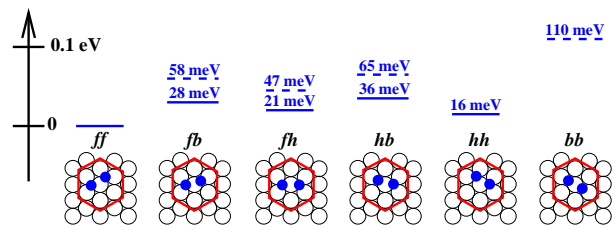}}
        \caption[4]{
        Illustration of the DFT analysis of the dimer intra-cell
        dynamics~\cite{Bogicevic98} observed on
        Cu(111). Solid (dashed) lines illustrate the variation in 
        the fully relaxed dimer-configuration energies
        calculated using the $4\times4$ ($2\times2$) $k$-point
        sampling.}
\protect\label{dimersch}
\end{figure}
From calculations
using $2\times2$ $k$-point sampling (Fig.\,\ref{dimersch}, dashed lines),
we predict an intra-cell dimer-energy variation that is in qualitative
agreement with the original Al-dimer study ~\cite{Bogicevic98}.
As in these previous theoretical calculations and consistent with
the experiments, we find that intra-cell rotation remains possible
even when low temperatures inhibit the net inter-cell diffusion,
for which we determine a concerted-sliding barrier $E_{\rm B,dif} 
\approx \hbox{190\,meV}$. Using the more accurate $4\times4$ $k$-point
sampling we predict an energy variation (Fig.\,\ref{dimersch}, solid lines), in
even better agreement with experimental observations. The
$4\times4$ $k$-point calculations establish a near degeneracy
between the $ff$, $fh$, and $hh$ dimer configurations
and a significantly reduced effective rotation barrier, i.\,e.
the energy of the $fb$ configuration.
The accuracy improvement obtained by using the $4\times4$ $k$-point
sampling in the dimer-configuration survey is selective and
correlated with the dimer-induced surface distortion as also
previously found in the Al-addimer study \cite{Bogicevic98}.
We attribute the experimental observation of 
dimer confinement to  
the rotation process, $ff \to fb \to fh$ (and back),
which has a lower barrier than the $ff \to fb \to 
fh \to hb \to hh$ rotation and the direct $ff\to bb\to hh$ 
transition.

Trimers were found to be always of fcc-fcc-fcc type and stable 
in the temperature range of $5-21$\,K.

\begin{table}
\caption{
        Site preferences, kinetic barriers, and stability of monomers, dimers, 
        and trimers on Cu(111) as observed in low-temperature STM and calculated
        in first-principles DFT.  
        \protect\label{rot.tab}
        }
\begin{tabular}{llr|c|c|}
      &         &                        & {\bf STM} & {\bf DFT}\\
\hline
\multicolumn{2}{l}{\bf Monomer:} & & & \\
              & preferred site      &    & fcc & fcc \\
              & diffusion barrier $E_B$ & (meV)   & $37\pm5$ & $50$ \\
              & attempt frequency  $\nu_0$ & (s$^{-1}$) & $5\times10^{13\pm1}$ & $1 \times 10^{12}$ \\ 
              & $\Delta E = E_{\rm hcp}-E_{\rm fcc}$ & (meV) & $4\leq\Delta E\leq8$ & $6$ \\
\hline
\multicolumn{2}{l}{\bf Dimer:} & & & \\
              & preferred site     &       & fcc-fcc & fcc-fcc \\
              & intra-cell-diffusion $E_B$ & (meV) & $18\pm3$ & $28$ \\
              & $E_{\rm fcc-hcp}-E_{\rm fcc-fcc}$ & (meV) & $1.3\pm0.5$ & $21$ \\
\hline
\multicolumn{2}{l}{\bf Row of adatoms:} & & & \\
              & preferred site          & & fcc & fcc \\ \hline
\end{tabular}
\label{SumTable}
\end{table}
In summary, we have shown that individual copper adatoms as well as adatoms in
dimers, trimers, and larger structures prefer fcc over hcp sites on Cu(111).
The 
experimentally observed difference in binding 
energy for the monomer is only about 6\,meV.
Moreover, in the case of copper dimers, we were able to 
observe a confined 
intra-cell metal dimer motion for a wide 
temperature range of $4-21$\,K, which at low temperatures 
is enabled by thermally assisted tunneling.
Even though 
the small energy differences inherent in 
this system may attain the limits of 
current DFT accuracy,
a detailed DFT analysis 
yields monomer and dimer energetics that are in very good agreement with the experiments (see Table \ref{SumTable}).

We thank Prof.\ G.\ Ehrlich and Prof.\ M.\ Scheffler 
for their very valuable comments, and acknowledge partial funding by the EU TMR
project ``AMMIST'', the EU IST-FET project 
``NICE'', the Deutsche Forschungsgemeinschaft Project No.\ RI 472/3-2, 
the Chalmers UNICC (for supercomputing), and the Swedish 
Foundation for Strategic Research via ATOMICS (PH).


\begin{thebibliography}{99}
\bibitem{Wang90} S. C. Wang and G. Ehrlich, Surf. Sci. {\bf 239}, 301 (1990).
\bibitem{Busse03}C. Busse {\it et al.}, Phys. Rev. Lett. {\bf 91}, 056103 (2003).
\bibitem{WangEhrlich89}S. C. Wang and G. Ehrlich, Phys. Rev. Lett. {\bf 62}, 2297 (1989); Surf. Sci. {\bf 224}, L997 (1989).
\bibitem{WangEhrlich9192}S. C. Wang and G. Ehrlich, Surf. Sci. {\bf 246}, 37 (1991); Phys. Rev. Lett. {\bf 68}, 1160 (1992).
\bibitem{Goelzhaeuser96}A. G\"olzh\"auser and G. Ehrlich, Phys. Rev. Lett. {\bf 77}, 1334 (1996).
\bibitem{Piveteau92} B. Piveteau, D. Spanjaard, and M. C. Desjonqu\`eres, Phys. Rev. B {\bf 46}, 7121 (1992).
\bibitem{Papadia96} S. Papadia, B. Piveteau, D. Spanjaard, and M. C. Desjonqu\`eres, Phys. Rev. B {\bf 54}, 14720 (1996).
\bibitem{StumpfScheffler}R. Stumpf and M. Scheffler, Phys. Rev. Lett. {\bf 72}, 254 (1994);
Phys. Rev. B {\bf 53}, 4958 (1996).
\bibitem{Feibelman92}P. J. Feibelman, Phys. Rev. Lett. {\bf 69}, 1568 (1992).
\bibitem{Kuerpick01}  U. K\"urpick, Phys. Rev. B, {\bf 64}, 075418 (2001).
\bibitem{Boisvert95} G. Boisvert, L. J. Lewis, M. J. Puska, and R. M. Nieminen, Phys. Rev. B {\bf 52}, 9078 (1995).
\bibitem{GiesenIbach}M. Giesen and H. Ibach, Surf. Sci. {\bf 529}, 135 (2003).
\bibitem{Bogicevic98}A. Bogicevic, P. Hyldgaard, G. Wahnstr{\"o}m, and B. I. Lundqvist, Phys. Rev. Lett. {\bf 81}, 172 (1998).
\bibitem{DacapoLB} DACAPO ({\tt http://www.fysik.dtu.dk/CAMPOS}) version 1.30 with extensions by L. Bengtsson using GGA(PW91).
\bibitem{PtPuzzle01} P. J. Feibelman {\it et al.},
J. Phys. Chem. B {\bf 105},  4018 (2001).
\bibitem{Repp00}J. Repp {\it et al.}, Phys. Rev. Lett. {\bf 85}, 2981 (2000).
\bibitem{Knorr02}N. Knorr {\it et al.}, Phys. Rev. B {\bf 65}, 115420 (2002).
\bibitem{Wulfhekel96}W. Wulfhekel {\it et al.}, Surf. Sci. {\bf 348}, 227 (1996).
\bibitem{Stoltze94}P. Stoltze, J. Phys.: Condens. Matter {\bf 6}, 9495 (1994).
\bibitem{EiglerSchweizer90}D. M. Eigler and E. K. Schweizer, Nature {\bf 334}, 524 (1990).
\bibitem{Meyer00}G. Meyer {\it et al.}, Single Molecules {\bf 1}, 79 (2000).
\bibitem{Heinrich02}A. J. Heinrich, C. P. Lutz, J. A. Gupta, and D. M. Eigler, Science {\bf 298}, 1381 (2002). 
\bibitem{TrimerRowDFT} For the row-of-adatom case
our GGA-DFT (3-by-1, five-layer unit cell, 4-by-12 
$k$-points, 30\,Ry cutoff) yields a 8.6\,meV/atom fcc preference.
\bibitem{monovibration} $\nu_0=1\times 10^{12}$\,s$^{-1}$ 
results from extensive finite-difference DFT calculations of the local fcc- and hcp-monomer 
dynamics as will be described in a forthcoming publication.
\bibitem{Bogicevic00}A. Bogicevic {\it et al.}, Phys. Rev. Lett. {\bf 85}, 1910 (2000).
\bibitem{HyldgaardPersson00}P. Hyldgaard and M. Persson, J. Phys.: Condens. Matter {\bf 12}, L13 (2000).
\bibitem{Fichthorn00} K. A. Fichthorn and M. Scheffler, Phys. Rev. Lett. {\bf 84}, 5371 (2000).
\bibitem{Goran} G. Wahnstr{\"o}m, in ``Interactions of Atoms 
and Molecules with Solid Surfaces'', V. Bortolani, N. H. March, 
and M. P. Tosi, eds. (Plenum Press, New York, 1990), p. 529.
\bibitem{LauhonHo}L. J. Lauhon and W. Ho, Phys. Rev. Lett. {\bf 85}, 4566 (2000); {\bf 89}, 079901 (2002).
\bibitem{HillWheeler}D. L. Hill and J. A. Wheeler, Phys Rev. {\bf 89}, 1102 (1953).
\end{thebibliography}
\end{document}